# AI 驱动的 6G 空口：技术应用场景与均衡设计方法

王晓云、韩双锋、刘志明、王启星、王江舟、易芝玲

**摘要**：为应对 6G 系统在容量、频谱效率、能量效率等关键性能指标上面临的严峻挑战，将人工智能（Artificial intelligence, AI）技术引入空口传输已成为重要的技术突破方向。然而，当前 6G 空口 AI 研究主要聚焦于设计高精度 AI 模型以提升通信能力，普遍忽视了工程实践中所需的算力、复杂度和空口资源等 AI 代价，对模型泛化能力和推理时延等关键 AI 质量指标也缺乏系统性考量。这种过度依赖算力资源追求通信能力提升的研究范式，难以支撑智能化网络的可持续发展。本文系统分析了 AI 在 6G 空口传输中的典型应用场景与关键技术挑战，涵盖单功能模块性能增强、多功能模块联合优化以及复杂数学问题低复杂度求解等重要领域。创新性地提出了综合考虑 AI 能力、质量和代价的三维联合优化设计准则，通过最大化多场景通信能力与综合代价的比值实现三角均衡，有效弥补了现有设计准则中质量和代价维度缺失的不足。通过多个设计实例验证了所提方法的有效性，并深入探讨了空口 AI 标准化面临的技术路径与挑战，为 6G 空口 AI 技术的理论研究与工程实践提供了重要参考。

**关键词**：6G 空口, 人工智能, 均衡设计, 频谱效率, 能量效率，AI 质量，AI 代价

# AI-driven 6G Air Interface: Technical Usage Scenarios and Balanced Design Methodology

Xiaoyun Wang, Shuangfeng Han, Zhiming Liu, Qixing Wang, Jiangzhou Wang, Chih-Lin I

Abstract: To address the severe challenges faced by 6G systems in key performance metrics such as capacity, spectral efficiency and energy efficiency, the introduction of Artificial Intelligence (AI) technology into air interface transmission has become a significant technological breakthrough. However, current industry researches on AI enabled 6G air interface primarily focus on designing high-precision AI models to enhance communication capabilities, often overlooking the AI costs associated with computational power, complexity, and air interface resources required in practical engineering applications. There is also a lack of systematic consideration of critical AI quality metrics such as model generalization capability and inference latency. This research paradigm, which excessively relies on computational resources to pursue communication performance improvements, is unsustainable for the development of intelligent networks. This paper systematically analyzes the typical application scenarios and key technical challenges of AI in 6G air interface transmission, covering important areas such as performance enhancement of single functional modules, joint optimization of multifunctional modules, and low-complexity solutions to complex mathematical problems. Innovatively, a three-dimensional joint optimization design criterion is proposed, which comprehensively considers AI capability, quality, and cost. By maximizing the ratio of multi-scenario communication capability to comprehensive cost, a triangular equilibrium is achieved, effectively addressing the lack of consideration for quality and cost dimensions in existing design criteria. The effectiveness of the proposed method is validated through multiple design examples, and the technical pathways and challenges for air interface AI standardization are thoroughly discussed. This provides significant references for the theoretical research and engineering practice of 6G air interface AI technology.

Indexed terms: 6G air interface, Artificial intelligence, balanced design, spectrum efficiency, energy efficiency, AI quality, AI costs

# 1. 引言

为应对未来6G无线通信系统在容量、峰值速率、用户体验速率、频谱效率（定义为速率与带宽的比值）以及能量效率（定义为速率与功率的比值）等关键性能指标方面的严峻挑战[1]，业界普遍考虑在5G技术基础上通过增加基站密度、天线数量、射频通道数、并行传输数据流数以及扩展带宽等方式来提升系统性能。然而，这种单纯依赖资源堆叠的方式不仅会显著增加系统复杂度、功耗和成本，还难以高效、绿色地满足6G网络的多样化需求[2]。近年来，人工智能（Artificial Intelligence, AI）技术的快速发展为解决这一问题提供了新的思路。AI技术凭借其强大的复杂问题建模能力、动态策略优化能力以及基于通用算力的低复杂度计算与快速迭代能力，成为突破6G空口传输设计瓶颈的关键技术手段之一。

AI技术已在空口传输多个技术应用场景取得显著进展，涵盖了AI赋能的编码、调制、多址接入、波形设计、多输入多输出（Multiple Input Multiple Output, MIMO）预编码、信道估计、信号检测、语义提取以及资源调度等功能模块的设计与优化。国际标准组织3GPP已启动物理层AI技术的标准化工作，为空口AI技术的实际应用奠定了重要基础。然而，当前国内外在空口AI领域的研究仍面临诸多开放性问题与挑战。

首先,空口 AI 应用场景不明确,以 AI 使能波束管理、用户定位、MIMO 信道状态信息获取等问题为核心的技术解决方案尚未具备标准化的共识基础,且包括运营商、基站设备提供商、终端及芯片供应商也缺少共赢的用例。其次,空口 AI 设计评估准则忽略了包括数据采集及训练成本等要素,缺少可持续发展能力。现有多数研究仅关注利用 AI 如何提升空口容量和频谱效率等传输能力,而忽略了数据采集、模型训练、模型分发等 AI 必须步骤增加的复杂度、资源开销和算力成本等代价。第三,如何测试空口 AI 在不同场景与不同系统参数下的泛化能力,尤其与 AI 质量密切相关的模型推理时间,在产业界仍未形成基准。目前,仅以空口 AI 传输能力最大化为单一目标的设计准则不具备可持续发展能力。

针对上述空口 AI 面临的问题与挑战,本文首先系统分析了 AI 技术在 6G 空口中的典型应用场景,包括空口传输单模块性能增强、多模块联合创新设计、复杂问题低复杂度求解等三个方面,探讨了 AI 技术的潜在价值与作用,并分析了存在的挑战。在此基础上,本文提出了一种面向空口 AI 能力-质量-代价均衡优化的设计评估方法,通过最大化多场景通信能力与综合代价的比值实现三角均衡,并结合具体设计示例,阐述了如何降低数据采集与数据集构建的开销、优化模型训练与推理复杂度,以及实现空口 AI 单模块与多模块的能力-质量-代价联合优化。最后,本文对 6G 空口 AI 传输以及相应的接入网架构标准化的技术路径和挑战进行了展望。

## 2. 6G 空口 AI 的典型技术应用场景

在经典信息论的框架下,传统信号处理算法能够有效解决理想信道条件下的收发机优化设计问题。例如,通过构建线性化的调制解调、编译码以及 MIMO 预编码等方法,可以显著提升系统性能。部分调制技术甚至能够使传输信号呈现近似高斯分布,从而在理论上逼近香农极限。然而,在某些实际技术应用场景中,传统非 AI 技术在下面三种情况下面临难以克服的挑战。一是对复杂的通信相关数学问题难以建模的情况,例如在复杂信道条件下对噪声和干扰难以建模、对信道估计与多用户 MIMO 预编码等多模块联合优化难以建模等;二是即使能够建模,却难以找到有效的求解方法,例如多小区的用户联合调度与分布式 MIMO 预编码;三是虽然能够建模并存在求解算法,但由于计算复杂度过高而难以实际应用,例如需要复杂高维矩阵求逆运算的 MIMO 预编码和 MIMO 检测等。针对

这些挑战，将 AI 技术引入相关应用场景，利用神经网络对复杂问题进行数学建模和求解，有望显著提升系统的容量、频谱效率和能量效率等关键性能指标。因此，AI 技术成为 6G 空口设计中的一种重要技术手段，为未来通信系统的优化提供了新的可能性。

6G 空口 AI 包括 AI 赋能的物理层、数据链路层（QoS 流处理、链路自适应、资源调度等）、网络层（移动性管理、网络节能、负载均衡、覆盖与容量优化等）。本文的分析中，我们聚焦在物理层传输技术，典型技术应用场景如表 1 所示。

第一类是空口单功能模块基于 AI 的增强，此类场景聚焦于在现有空口传输协议的框架下，通过引入 AI 技术对单一功能模块进行优化，旨在显著提升 6G 空口的频谱效率、能量效率及传输可靠性等关键性能指标。例如，利用深度学习优化信道估计、预测、压缩反馈的精度，用轻量化模型对器件非线性进行补偿等，从而在理论极限范围内进一步挖掘系统潜力。

第二类是多功能模块的联合创新设计，此类场景致力于基于 AI 的多模块联合优化甚至全链路智能化设计，旨在突破传统模块化设计的局限性，推动空口传输领域实现颠覆性的技术创新。例如，通过 AI 技术实现 TDD 系统中上行信道估计、预测、MIMO 预编码联合设计，下行导频估计、MIMO 检测、收发器件非线性补偿等联合设计，以及发射与接收链路基于 AI 的联合设计。此类研究不仅提升了系统整体性能，还为未来通信系统的架构设计提供了新的科学思路。

第三类应用聚焦于空口传输中复杂数学问题的低复杂度求解。这类场景主要针对空口传输过程中产生的高维、非线性复杂数学问题，虽然传统方法能够求解，但往往面临计算复杂度过高的挑战。通过引入 AI 技术，利用其高效的推理能力，可以在有限算力约束下实现更优的算法性能，从而有效降低计算复杂度，提升系统整体效率。例如，利用神经网络近似高维矩阵求逆运算，降低 MIMO 检测算法的计算复杂度。此类研究为复杂通信问题的实际工程化应用提供了可行的科学解决方案。

在三类应用场景中，业界有大量的研究成果，表 1 仅列出了一些代表性工作。

表 1 6G 空口 AI 的典型技术应用场景

| 应用类别 | 具体用例 | 目标 | 示例 |
|---|---|---|---|
| 单功能模块的性能 | 基于卷积神经网络的信道估计 | 提高信道估计精度 | 文献[3] |
| | 基于长短时记忆网络的波束预测 | 提高波束选择正确率 | 文献[4] |

| 增强 | 基于 Transformer 网络的信道预测 | 提高信道预测精度 | 文献[5] |
|---|---|---|---|
| | 基于 AI 的信道压缩与反馈 | 提高信道获取精度 | 文献[6] |
| | 基于 AI 的器件非线性补偿 | 提高频谱效率 | 文献[7] |
| 多功能模块的联合创新设计 | 基于一个神经网络的上行信道估计、预测、MIMO 预编码的多模块建模 | 降低推理复杂度和推理时延 | 文献[9] |
| | 基于 AI 的通信收发全链路设计 | 提高通信效率，提高通信系统迭代速度 | 文献[10] |
| 复杂问题的低复杂度求解 | 基于 AI 的和速率最大的 MIMO 预编码优化 | 降低推理复杂度 | 文献[12] |
| | 基于 AI 的频谱效率最大的预编码和功控策略 | 降低推理复杂度 | 文献[13] |

2.1 AI 助力单功能模块的性能增强

AI 助力单功能模块的性能增强方面，业界研究了通信链路所有模块，包括 AI 赋能的编译码、调制解调、波形、多址、MIMO 信道估计/预测/压缩反馈、MIMO 预编码/检查、波束管理、定位增强、非正交导频、器件非理想特征的补偿等。从对系统频谱效率提升的角度来看，大规模 MIMO 的信道获取技术以及收发机器件的非线性补偿是非常重要的技术应用场景，下面将重点分析。

1）大规模 MIMO 技术作为 6G 通信系统实现高频谱效率与能量效率的关键技术路径，其性能优化面临显著挑战。在系统实际部署中，上下行精确信道信息的获取需要消耗大量导频资源，特别是在频分双工（Frequency Division Duplex, FDD）系统中，终端反馈过程会带来额外的时频资源开销。虽然增加导频密度有助于提升信道估计精度、优化下行波束赋形性能并改善用户数据信号解调质量，但这也意味着需要占用更多的时频资源。研究表明，以下行吞吐量最大化为优化目标时，下行导频密度存在最优配置方案。传统信道估计与预测算法不仅受限于链路时偏、频偏及干扰等难以精确建模的非理想因素，还面临着计算复杂度高的瓶颈问题。

人工智能技术的引入为上述挑战提供了创新解决方案。AI 赋能的信道估计与预测方法能够在更低导频密度条件下实现精确信道信息获取，同时基于 AI 的信道压缩与恢复技术可显著降低反馈开销，使基站获得高精度的信道状态信息，从而有效提升大规模 MIMO 系统的频谱效率。具体而言，文献[3]创新性地提出了一种基于卷积神经网络（Convolutional Neural Network, CNN）的信道估计网络架构，通过将信道矩阵建模为二维图像，充分挖掘其在频域与空域的相关特性。

仿真实验表明，该方案较传统最小二乘（Least Square, LS）估计方法可提升 10dB 的估计精度，且性能逼近最小均方误差（Minimum Mean Square Error, MMSE）估计器。针对移动场景下的波束预测难题，文献[4]设计了一种基于长短时记忆网络（Long Short-Term Memory, LSTM）的波束预测模型，并创新性地引入概率机制以提升波束选择的准确率，仿真结果验证了该网络在高速移动场景下仍能保持准最优的波束预测性能。为进一步提升信道预测的时间跨度，文献[5]提出了一种基于 Transformer 架构的信道预测网络，充分利用 Transformer 的并行预测优势，实现了对未来多个时隙的精确信道预测，其长时预测性能显著优于传统算法。此外，将 AI 技术应用于大规模 MIMO 系统的信道压缩与反馈过程，已成为提升 FDD 系统基站侧信道获取精度的有效手段[6]，为未来 6G 通信系统的实际部署提供了重要技术支撑。

2）由于收发信道的硬件器件存在复杂的非线性特性，接收机难以获取精确的端到端信道信息，导致自适应传输技术的实现面临显著挑战。在 6G 通信系统中，随着系统带宽的进一步扩展和动态性的显著增强，信道环境的非线性效应和时变特性变得更加突出。传统解决方案通常采用数字预失真（Digital Pre-Distortion, DPD）技术对发射信号进行预处理，以补偿功率放大器的非线性效应，从而保证发射信号的线性度。然而，这种方法需要在发射机侧部署复杂的数字预失真模块，其较高的实现复杂度使其难以在终端侧广泛应用，因此无法有效补偿上行链路的非线性失真。此外，仅基于发射端的非线性补偿难以对发射机与接收机之间全链路的非线性特性进行精确建模和完全补偿。

相比之下，在接收端对全链路的非线性特性进行建模并实施补偿是一种更为理想的技术路径。然而，传统非 AI 方法难以对这种复杂的非线性规律进行有效建模。文献[7]通过仿真和样机验证表明，AI 技术能够通过低复杂度的神经网络实时在线学习收发链路的非理想特性，并在接收端实现精确补偿，从而显著降低发射机（如终端侧）的实现复杂度。实验结果表明，在相同的误差矢量幅度（Error Vector Magnitude, EVM）指标下，基于 AI 的非线性补偿技术支持更高的调制阶数，从而有效提升通信系统的频谱效率。在通信链路容量逐渐逼近香农极限的背景下，AI 驱动的非线性补偿技术已成为提升系统实际频谱效率的关键手段，为 6G 通信系统的性能优化提供了新的技术方向。

基于上述分析，AI 技术在大规模 MIMO 信道获取与器件非线性补偿方面展现出显著优势，其性能超越了传统非 AI 技术，成为 6G 系统提升频谱效率的核心技术手段。然而，现有的基于 AI 的单模块增强技术仍面临诸多亟待解决的开放性问题：1）在 AI 模型设计方面，如何充分利用多维度的信道信息仍需深入探索。在信道估计、预测、压缩反馈等应用场景中，信道的空域、时域、频域相关性以及上下行信道的部分频域相关性均可被有效挖掘，以实现更优的信道获取性能。现有方法往往局限于利用部分信息进行 AI 模型设计，例如在信道压缩场景中，通常仅基于空频域下行信道矩阵或特征向量进行压缩与反馈，未能充分挖掘信道在多个维度的相关性特征，限制了性能的进一步提升。2）针对具有相关性的应用场景，AI 模型的独立设计与通用化设计之间的权衡需要深入研究。具体而言，对于 DMRS（解调参考信号）、Sounding RS（上行探测参考信号）、CSI-RS（信道状态信息参考信号）等不同参考信号的估计算法，是否可以设计统一的 AI 模型架构？此外，大规模 MIMO 系统中的信道估计、预测、压缩反馈等不同任务虽然都需要基于 AI 模型挖掘信道特征，但这些任务是应该分别设计专用模型，还是可以采用联合优化的设计思路？如果采用联合设计，其具体方法学和技术路径又该如何构建？这些问题都需要在未来的研究中得到系统性解答。

2.2 AI 助力通信链路多功能模块的联合创新设计

经过五代移动通信系统的持续演进，物理层关键技术，包括信道编码、调制、波形设计、MIMO 以及正交频分复用（Orthogonal Frequency Division Multiplexing, OFDM）[8]等已趋于成熟。然而，现有通信系统的各个功能模块通常采用独立设计和优化的方式，基于不同的假设和目标，导致系统存在累积误差和分段处理效率低下等问题。由于多模块联合优化的数学问题难以建模和求解，传统方法难以实现全局最优设计。

AI 技术的引入为解决这一问题提供了新的思路。以时分双工（Time Division Duplex, TDD）系统为例，通过 AI 技术可以将上行信道估计、信道预测、MIMO 预编码等多个相关功能模块集成到一个统一的神经网络框架中[9]，将复杂的多模块联合优化问题转化为高效的数据拟合或回归问题，从而获得接近全局最优的解决方案。这种端到端的设计策略能够隐式地实现信道估计、预测与预编码的联合优化，不仅显著降低了推理复杂度，还减少了处理时延。此外，AI 赋能的通

信收发全链路设计[10]也具有重要的研究价值，其潜力在于突破现有通信功能模块的专利壁垒和标准化限制，推动通信系统向智能化方向快速迭代。这些创新设计思路为通信系统优化开辟了新的技术路径，而传统非 AI 技术由于实现难度大、可行性低，难以达到类似的效果。AI 技术的引入不仅提升了系统性能，也为未来通信系统的智能化发展奠定了重要基础。

尽管从理论上看，设计神经网络以替代通信链路中的多模块甚至全链路功能具有可行性，但相关研究仍处于初步探索阶段，大量开放性问题亟待解决。例如，在通信链路中，哪些模块适合联合设计以实现最优性能？哪些模块更适合独立优化？又有哪些模块无需依赖 AI 技术进行增强？此外，端到端通信链路的智能化设计是否代表了未来的技术发展方向？这些问题尚未得到充分研究和明确答案。

2.3 AI 助力复杂问题的低复杂度求解

在 6G 通信系统中，多小区多用户场景下的资源调度、预编码设计、功率控制等优化问题通常具有非凸特性，且优化变量涉及高维空间，难以直接获得全局最优解。现有研究大多基于经典优化理论和非凸转凸技术进行求解，但这类方法普遍存在计算复杂度高、易陷入局部最优解的问题。特别是在快时变信道环境中，由于信道状态信息的快速变化，迭代优化过程的求解时间较长，导致所得最优解的时效性难以保证。此外，6G 系统的大规模天线阵列、射频通道以及并行数据流传输等特性，使得传统信道估计和 MIMO 检测算法中涉及的高维矩阵运算（如矩阵求逆、奇异值分解等）的计算复杂度和时间开销急剧增加，难以满足实际系统的实时性需求。以多用户 MIMO 系统中的和速率最大化预编码优化为例，该问题本质上是非凸优化问题，通常采用加权最小均方误差（Weighted Minimum Mean-Square Error, WMMSE）算法求解。然而，WMMSE 算法需要通过多次迭代获得数值解，其计算复杂度随天线数的立方和用户数的平方增长。由于最优预编码依赖于环境参数（如信道系数），基于特定信道矩阵计算得到的预编码仅在信道相干时间内（通常为毫秒级）有效，这进一步加剧了实时资源分配的难度[11]。

AI 技术为解决上述挑战提供了新的思路。通过数据驱动或模型驱动的方式，AI 能够直接学习输入信息与潜在解决方案之间的映射关系，从而快速给出问题的近似解或精确解。现有研究表明，当采用卷积神经网络（CNN）学习和速率最大化的多用户 MIMO 预编码时，在天线数 N 为 8~256、用户数 K 为 4~120 的

场景下,其推理复杂度仅为 WMMSE 算法的 1/2~1/100[12]。此外,用于 MIMO 预编码的神经网络结构还可以进一步优化以降低复杂度。无线通信领域中存在大量成熟的数学模型,例如空口数据速率与传输资源及信道参数之间的关系。近年来,越来越多的研究聚焦于模型驱动的深度神经网络设计,其核心思想是利用特定问题的最优解结构或数值算法的显式/迭代框架,将传统算法中计算复杂度较高或难以确定的模块替换为神经网络。例如,在学习频谱效率最大化的预编码和功率控制策略时,可以通过深度展开 WMMSE 算法,利用神经网络替代矩阵求逆或迭代方向与步长的计算[13]。因此,AI 技术在复杂问题求解中的应用不仅能够显著降低计算复杂度,还能有效克服传统非 AI 方法在实时性和计算资源需求方面的局限性,为 6G 通信系统的优化设计提供了全新的技术路径。

综上所述,AI 技术有望在空口传输单模块性能增强、多模块联合创新设计、复杂问题低复杂度求解等一些典型技术应用场景中发挥重要作用,助力实现 6G 系统的需求指标。但是,业界对空口 AI 的设计评估准则还没有达成共识,很多研究还是侧重于大算力前提下的通信能力最大化,对引入 AI 技术后所带来的数据、算力、复杂度等代价以及 AI 模型的泛化性、推理时延等缺乏系统性思考和优化。作为 5G 和 6G 通信系统智能化演进的方向指引,科学完善的空口 AI 设计评估准则不仅是技术创新的基础保障,更是通信产业可持续发展的先决条件。下面我们将探讨面向空口 AI 能力-质量-代价三角均衡优化的设计准则与一些示例。

## 3. 能力-质量-代价均衡优化的空口 AI 技术设计准则

空口 AI 技术对于通信系统来说是一个全新的概念,设计评估理念方面需要有全新的思考和系统创新。首先,AI 模型要满足空口 AI 的通信能力指标需求,例如频谱效率和能量效率,这就要求 AI 模型的推理准确度、推理时延、复杂度要满足需求。此外,随着室内、室外、高铁、体育馆等典型部署场景的变化,通信信道特征和 AI 模型的输入输出参数会相应发生变化,AI 模型在不同场景中的泛化能力也要满足需求。其次,AI 赋能空口性能提升所需付出的代价也要综合考虑,包括数据集构建、模型训练/再训练、模型分发与部署等相关的数据、算力、复杂度、空口时频资源等代价。

### 3.1 设计评估准则

为了更加全面的评估智能空口的性能，我们提出空口 AI 能力-质量-代价三角形均衡设计评估模型。如图 1 所示，空口 AI 能力包括引入 AI 技术后频谱效率、能量效率、系统容量等空口通信能力指标；空口 AI 代价代表为了达到预期的空口传输能力所付出的 AI 代价，包括 AI 模型相关的数据采集、数据集构建、模型训练、模型分发等所需的算力、复杂度、空口时频资源等开销；空口 AI 质量是 AI 模型的质量，包括推理准确度、推理时延、不同场景下的泛化性等。

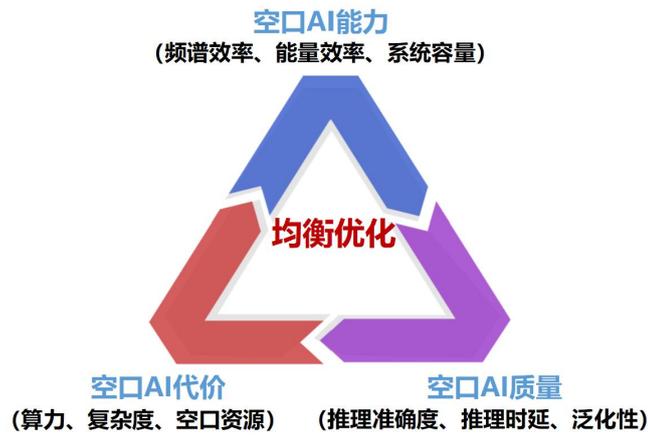

图 1 空口 AI 能力-质量-代价三角形均衡设计评估模型

高质量的 AI 模型是提升频谱效率等能力的基础。然而，推理准确度和泛化能力的提升往往需要更丰富的训练数据、更高的训练与推理复杂度以及更多的空口时频资源开销，这会导致代价的显著增加。此外，空口 AI 的推理过程需融入空口资源调度和数据传输流程，其推理时延必须控制在调度周期（例如 1ms）之内。推理时延可通过推理复杂度与算力的比值计算得出，在算力受限的情况下，降低推理时延通常需要简化 AI 模型，但这可能以牺牲推理准确度和泛化能力为代价。因此，在空口 AI 质量优化中，推理准确度/泛化性与推理时延之间存在固有的矛盾。

空口 AI 能力指标之间也存在矛盾，例如频谱效率与能量效率。由于基站需要周期性发送下行广播信号和参考信号，系统的静态功耗不为零，因此频谱效率与能量效率之间的关系极为复杂。当频谱效率达到一定阈值后，其进一步增长往往会导致能量效率的下降[14]。

因此，空口 AI 能力、质量与代价的联合优化是一个高度复杂的多目标优化问题。为了简化问题，我们对其进行了合理的拆解。首先，AI 模型的推理准确

度直接决定了空口 AI 能力（例如，基于 AI 模型的 MIMO 信道推理准确度直接影响 MIMO 容量），因此我们将对推理准确度的要求近似等价为空口 AI 能力的要求。其次，推理时延是一个硬性约束，所有可选的 AI 模型都必须满足时延要求，这可以通过优化代价（算力与模型推理复杂度）来实现。因此，在满足推理时延的前提下，单部署场景中的能力-质量-代价均衡优化可以通过最大化能力与代价的比值来实现。

当考虑多部署场景时，AI 质量中的泛化性成为关键指标。泛化性的本质是模型在未见数据上的表现，通常可以通过偏差（衡量模型推理值与真实值之间的差异）和方差（衡量模型对训练数据微小变化的敏感性）来评估。在本文中，我们采用同一空口 AI 模型在不同部署场景中的性能偏差或方差作为泛化能力的度量。随着场景的变化，AI 模型可能需要进行更新（如再训练）以满足泛化能力的要求，这会导致 AI 质量（推理准确度和时延）、代价以及空口能力的变化。如前所述，单部署场景中的均衡优化可通过最大化能力与代价的比值实现；而在多部署场景中，在满足泛化能力的前提下，能力-质量-代价的均衡优化可通过最大化多场景中能力与代价的综合比值（M）来实现。例如，当空口 AI 能力为频谱效率时，M 的一种表达式如下：

$$M=\frac{1}{N}\sum_{i=1}^{N}\frac{w_i\,SE_i}{C_i} \qquad (1)$$

其中，N 是场景数，$SE_i$ 是第 i 个场景中 AI 技术的频谱效率相比非 AI 技术频谱效率的增益（倍数），$w_i$ 是场景的加权，$C_i$ 是第 i 个场景中 AI 技术的综合代价（包含算力、复杂度、空口资源等）与非 AI 技术综合代价的比值。综合代价的定义可以有不同方式，包含综合成本（计算多种代价对应的总成本）以及多种代价的加权和等。

6G 不同应用场景（机器控制、全息通信、超能交通、孪生城市、感知类业务、智能交互等）中的时延、移动性、频谱效率等需求指标是不同的。针对不同的应用场景，我们都可以通过公式（1）来判断 AI 在多个部署场景（例如室内、室外、高铁、体育馆等）中发挥的作用和价值。

M 最大化的设计准则适用于空口 AI 单模块设计，例如信道估计、预测、MIMO 预编码、MIMO 检测等。该准则同样适用于一个 AI 模型实现多个串行模块功能的情况，例如 TDD 系统中的上行信道估计、信道预测、MIMO 预编码等功能。

如果空口 AI 有多个能力目标，比如既有频谱效率又有能量效率目标，就需要设计一个综合的能力指标，例如频谱效率和能量效率的加权。由于系统的能量效率和频谱效率单位不同，无法直接相加，所以也可采用增益（倍数）进行相加的方式，如下式所示。

$$M=\frac{1}{N}\sum_{i=1}^{N}\frac{w_i(EE_i+SE_i)}{C_i} \quad (2)$$

其中，$EE_i$ 和 $SE_i$ 是第 i 个场景 AI 技术带来的能量效率和频谱效率各自的增益（倍数）。值得注意的是，空口 AI 相关的数据采集、模型训练、推理等所需的功率要纳入计算能量效率时系统的总体功耗里面。

### 3.2 设计示例

为了在能力、质量和代价之间实现联合优化，我们提出两种策略：第一种是在确保空口 AI 能力和质量满足需求的前提下，通过降低空口 AI 的代价来提升 M；第二种是在空口 AI 能力满足需求的基础上，通过代价和质量的联合优化来提升 M。

针对第一种方法，我们可以在 AI 模型生命周期的每个环节采取有效措施以降低代价。具体而言：1）在数据采集阶段，可以在少量实测数据的基础上进行数据增强，从而减少数据采集的开销。当需要反馈采集数据时，可以仅反馈实测数据的特征，以降低反馈开销。2）在模型训练阶段，采用更先进的 AI 模型和训练方法，以降低训练复杂度。同时，使用特征更全面的数据集来提升模型的泛化性，减少再训练的需求。3）在模型再训练阶段，采用低训练开销的在线训练方法，降低模型再训练的复杂度。通过轻量化的迁移学习或元学习，提升 AI 模型对通信场景和参数的泛化能力，从而降低再训练的复杂度。4）在模型部署和推理阶段，通过低精度量化降低推理的复杂度，并利用模型蒸馏和剪枝技术减少模型的参数规模。我们在 3.2.1 节中提供了具体的降低代价的设计示例。

对于第二种方法，关键在于确保模型推理的准确度满足通信能力提升的要求，同时推理时延要符合空口用例的时延要求。在此基础上，主要挑战是如何在提升模型泛化性的同时，不增加甚至降低代价。这需要更加巧妙的 AI 模型设计，具备良好的泛化性、低训练开销、低再训练开销、低训练数据开销和低推理复杂度。在大语言模型设计方面，Deepseek 已经为业界提供了三角均衡优化的优秀示范。在空口 AI 的均衡设计方面，我们在 3.2.2、3.2.3 和 3.2.4 三节中分别给出了单模

块均衡优化、多模块均衡优化以及综合考虑频谱效率和能量效率指标的单模块均衡优化的示例。空口 AI 的均衡优化仍需业界进一步深入研究和探索。

### 3.2.1 降低空口 AI 代价示例

#### 3.2.1.1 降低数据采集和数据集构建的开销

构建基于少量实测数据的低开销数据集对于提升 M 至关重要，主要原因如下：1）通信场景和信道环境具有高度的多样性和复杂性，仅依赖仿真数据难以与实际环境完全匹配，从而影响 AI 模型的性能表现。2）获取实测数据集涉及数据测量、存储和反馈等环节，开销巨大（例如终端测量下行信道数据并反馈给基站），在实际应用中面临较大挑战。为降低实测数据的获取成本，可以在少量实测数据的基础上进行数据增强。我们在前期研究中，基于实测信道数据提取了信道在多径时延、角度、功率等方面的统计特征，并通过仿真平台生成了与多种实际信道环境高度吻合的信道数据集。仿真结果表明，该方法能够将实际数据采集开销降低 80%以上，同时将数据反馈开销减少 95%以上[15]。

#### 3.2.1.2 降低模型训练/再训练和推理复杂度

在 AI 领域，已有多种通过低复杂度神经网络设计来降低模型训练和推理复杂度的方法，这些方法能够显著提升 M 值。例如，采用低精度量化技术可以降低推理复杂度，而剪枝技术则能够有效减少模型的参数规模，从而提升整体效率。

此外，随着通信场景和系统参数的变化，AI 模型的再训练往往伴随着较高的开销。因此，降低再训练复杂度也是提升 M 值的关键手段之一。具体而言，有两种主要方式：一是采用低训练开销的在线训练方法。例如，在接收端通过神经网络对发射与接收器件的非线性特征进行补偿时，由于非线性特征变化较为缓慢，可以选择轻量化的 AI 模型，并基于极小量的训练数据进行在线训练[7]。二是通过轻量化的迁移学习或元学习技术，提升 AI 模型对通信场景和参数的泛化能力。此外，多任务学习可以在不同场景下选择最优的 AI 模型，从而避免模型的再训练[16]。如果希望在降低推理复杂度的同时提升模型的泛化性，可以采用低复杂度的图神经网络设计来实现[17]。这些方法为降低开销和提升性能提供了有效的解决方案。

### 3.2.2 空口 AI 单模块能力-质量-代价联合优化示例

以 MIMO 信道压缩反馈为例,我们针对 AI 算法在不同场景中泛化性不足的问题,提出了一种综合考虑终端和基站算力、基站与终端模型交互开销、模型在线训练开销以及多场景泛化性能的单编码器多解码器 AI 架构[16]。在该架构中,终端侧的编码器在不同通信场景中保持不变,其训练数据集涵盖所有场景的信道数据;而解码器则根据具体场景进行针对性设计(训练数据集仅包含单一场景的信道数据),并能够根据场景动态切换。这一设计显著降低了对终端算力和模型存储的需求,减少了基站与终端之间的信令开销,同时提升了 AI 算法在多场景下的性能表现。与基于多场景数据集单独训练的多编码器多解码器架构相比,所提架构在仅需 1/5 编码器存储开销的情况下,实现了相同的信道获取精度。这种 AI 模型设计结合前文所述的低开销信道数据集构建方法,可以很高的实现能力-质量-代价的联合优化。

### 3.2.3 空口 AI 多模块能力-质量-代价联合优化示例

以传统 TDD 系统的大规模 MIMO 信道获取和预编码为例,当天线端口数从 32 增加到 128 时,相应的导频开销占比会增至原来的 4 倍,这对系统频谱效率的提升极为不利。如果为了保持导频总开销不变而降低每个端口的导频密度,则会导致信道估计误差增加,进而使 MIMO 预编码性能下降。因此,当信道估计和 MIMO 预编码两个模块独立设计时,若按照固定导频密度进行配置,天线端口数的增加不仅无法带来频谱效率的等比例提升,反而可能导致频谱效率的恶化,造成严重的资源浪费。

在 TDD 系统中,可以通过 AI 模型实现上行信道估计、信道预测以及下行 MIMO 预编码等多个模块的联合设计。具体而言,神经网络以多用户信道信息作为输入,输出未来某一时刻每个用户的 MIMO 预编码矩阵,设计目标为最大化多用户和速率。与采用独立 AI 模型分别实现信道估计、预测和 MIMO 预编码的传统方式相比,多模块联合 AI 设计能够同时降低空口 AI 代价、提升能力并改善质量。主要原因如下:1)训练数据的采集开销大幅减少,模型训练和推理的复杂度显著下降,从而有效降低了空口 AI 的代价;2)与得益于 AI 技术在识别干扰、噪声特征以及信道相关性等方面的强大能力,即使在低导频开销下,也能实现精确的信道估计和预测。随着天线端口数的增加,导频开销不会等比例增长,甚至可能下降,从而显著提升了空口 AI 的能力;3)随着端口数的增加,传统预

编码算法所需的矩阵运算复杂度显著上升，而在算力受限的条件下，AI 驱动的预编码算法能够在低复杂度的前提下实现比传统算法更高的系统容量，从而进一步提升了空口 AI 的能力；4）由于 MIMO 预编码复杂度的降低，推理时延也得以减少，从而有效提升了空口 AI 的质量。这些优势共同推动了方案 M 值的提升。

我们提出了一种基于图神经网络（Graph Neural Network, GNN）的上行信道估计、预测和下行预编码多模块联合设计方案。与多种其他设计方案（包括全非 AI 设计、非 AI 设计的信道估计和预编码结合 AI 预测、非 AI 设计的信道估计结合 AI 预测和预编码等）相比，该方案在多个方面表现出显著优势：更高的下行吞吐量、更低的上行 sounding 参考信号开销、对信噪比和用户数等参数更强的泛化性，以及更低的计算复杂度[9]。根据 3.1 节提出的评估准则，该方案是一种综合性能优异的空口 AI 技术，优于论文中其他多种技术方案。结合低开销信道数据集构建，这种多模块联合设计方法可以更好的实现能力-质量-代价的联合优化。

3.2.4 空口 AI 单模块能力（综合效率）-质量-代价联合优化示例

在 5G 之前的移动通信系统中，提升频谱效率始终是核心目标。然而，5G 系统在设计时虽然引入了能量效率指标，却依然面临高功耗的挑战。这一问题的根源主要在于：首先，高频段通信和大规模 MIMO 技术的广泛应用显著增加了基站功耗；其次，网络密度的提升导致基站数量大幅增加，进一步推高了整体能耗；此外，5G 设备在非高峰时段的能耗优化机制尚未完善，节能技术的应用效率有待提升。这些因素共同导致了 5G 系统的能效表现未能达到预期水平。

在 6G 空口设计中，AI 技术的引入为频谱效率的提升带来了新的机遇，但同时也需警惕 AI 算力需求对能量效率的潜在影响。因此，兼顾频谱效率与能量效率的综合效率优化成为 6G 系统设计的关键课题。基于公式（2），我们在文献[18]中针对大规模 MIMO 信道压缩场景进行了优化设计。仿真结果表明，随着 AI 模型复杂度的提升，系统频谱效率的增益呈现边际递减趋势，而能量效率的衰减速率则显著加快。基于能力-质量-代价均衡优化框架，我们能够从多种 AI 模型中选择最优方案，在频谱效率、能量效率、模型复杂度、模型泛化性等多指标之间实现最佳平衡。

## 4. 6G 空口 AI 标准化技术路径和挑战分析

除了面向 6G 空口 AI 的前沿研究，产业界也在积极推动 5G 接入网和终端在现有标准框架下的智能化升级。AI-RAN 联盟在 5G 基础上聚焦三大研究和创新领域：利用 AI 提升接入网能力提高频谱效率，整合 AI 与接入网流程以更有效地使用基础设施，通过接入网在网络边缘部署 AI 服务以提高运营效率并为移动用户提供新服务。终端厂商也在加速 AI 与 5G 的融合，高通等企业推出了内嵌 AI 能力的 5G 芯片，支持运行更多 AI 专用的 5G 算法，不仅提升了连接体验，还拓展了多样化的应用场景。运营商在 AI 赋能 5G 接入网方面也取得了显著成果，例如，中国移动 AI 赋能的基站节能技术经现网实测可降低功耗达 9%，而 AI 赋能的业务感知准确率更是超过 95%。

在 5G 网络智能化运维方面，通信网络大模型发挥着越来越重要的作用。中国移动的网络大模型能够高效处理复杂的网络数据，提供意图智能识别、智能调度编排和自主优化决策能力，推动了"AI+网络"的深度融合创新。通过实现自动端到端闭环，该模型支持一线极简作业和无人干预的优化，覆盖了无线自闭环优化、基站隐患识别等多个场景，显著提升了网络运维的智能化和自动化水平。

3GPP 即将启动 6G 标准化工作，空口 AI 技术（涵盖物理层、数据链路层和高层）必将成为焦点。然而，近几年逐步落地的 5G 网络智能化技术对 6G 空口 AI 的标准化形成了一定的压力。与 5G 智能化技术相比，需要更加透彻地分析 6G 空口 AI 用例的必要性和技术优势，以确保其在实际应用中的价值。

6G 空口 AI 传输技术以及相应接入网架构标准化的技术路径和挑战分析如下：

1）空口 AI 传输技术标准化：技术路径方面，应顺序开展物理层单模块用例、多模块用例、全链路设计的标准化工作。设计评估准则是标准化讨论的重要基础，本文所提的三角均衡方法可以作为评估参考。目前标准化的第一个挑战是均衡评估方法还需深化，例如如何把算力、复杂度、时频资源等不同维度的代价进行综合需要进一步明确并达成共识。此外，如果既要满足频谱效率又要满足能量效率指标，空口 AI 的设计会更加复杂，尤其是 AI 相关的数据采集、模型训练、推理等与功耗的关系需要精确建模。第二个挑战是业界对多模块和全链路 AI 模型的研究还不深入，尚无法从三角均衡的角度判断怎样的路径是更科学的，是基于

AI 的单模块增强？多模块增强？还是全链路的 AI 设计？此外，由于无线信道的高动态性，AI 空口模型基于少量最新信道样本的快速训练非常重要，面向多模块和全链路优化的轻量化强泛化的新型 AI 模型有很高的研究价值。综合来看，空口 AI 传输技术的标准化将会是一个较长期过程。

2）接入网架构标准化：为了实时、灵活地支持不同空口 AI 算法的部署，需要设计高效的接入网架构[19]，以支持空口 AI 所需的数据、算力以及 AI 功能与协议栈的联合优化，从而实现 AI 的内生融合。在技术路径方面，标准化的初级阶段需要重点关注如何将 AI 功能或逻辑实体引入接入网架构，以满足 AI 模型在数据采集、模型训练、部署和推理等全生命周期管理中的需求。特别是对于实时性要求较高的空口 AI 应用场景（如信道估计、MIMO 预编码等），推理结果需要直接输入基站的调度器，这就要求 AI 功能或逻辑实体与现有基站协议栈之间建立高效的接口。然而，这一过程面临的主要挑战在于，如果网络需要支持第三方 AI 算法，相关接口必须实现标准化，而不仅仅是设备厂商内部的私有接口。这一标准化工作将面临较大的阻力，尤其是在多方利益协调和技术兼容性方面。

面向未来更加多样化的部署场景和更具挑战性的性能指标，空口 AI 的数据采集、模型训练与部署机制也需要进行相应的优化和调整，构建智能体赋能的场景与业务自适应的自演进空口 AI 体系化技术，具有重要的划时代意义。因此，在标准化的高级阶段，需要引入接入网智能体负责接入网的智能化工作[20]，涵盖数据面和控制面的物理层及高层协议的智能按需优化。具体而言，智能体需要具备对接入网任务的理解与分解能力，能够动态调用和更新 AI 模型库，以实现网络性能的持续优化。在这一过程中，通用大语言模型（如 Deepseek）可以发挥重要作用，通过智能体调用，帮助更准确地理解和分解任务。此外，大语言模型还可以助力空口 AI 数据集的构建和模型蒸馏等工作，从而提升空口 AI 模型的质量，同时降低其开发和部署成本。特别是在全链路 AI 模型的设计中，可以参考 Deepseek 等大模型在模型架构和训练方法上的创新思路，进一步提升空口 AI 的性能和效率。

然而，标准化高级阶段也面临诸多挑战，主要包括智能体基于任务的高效工作机制、不同智能体之间的协同机制、空口 AI 小模型的自动生成技术，以及大语言模型与空口 AI 小模型的协同优化等。这些问题的解决将为空口 AI 技术的广泛应用奠定坚实基础，并推动无线通信网络向智能化、自适应化方向迈进。

## 5. 总结

AI 技术与空口传输技术的融合有望显著提升空口传输能力，是 6G 空口研究的重要方向。业界在 6G 空口 AI 的研究聚焦在空口 AI 能力的提升，往往忽略 AI 代价和质量。本文分析了 AI 引入 6G 空口的典型技术应用场景，提出了空口 AI 能力-质量-代价三角形均衡设计准则，给出了若干设计示例，并探讨了空口 AI 传输以及相应的接入网架构标准化的技术路径和挑战。